\documentclass{ws-p10x7}

\def\pinn{$K^{+} \rightarrow \pi^{+} \nu \bar{\nu}$ }
\def\kmu2g{$K^{+} \rightarrow \mu^{+} \nu \gamma$ }
\def\kpi2g{$K^{+} \rightarrow \pi^{+} \pi^{0} \gamma$ }

\begin{document}

\title{Recent results from the BNL E787 experiment}

\author{Takeshi K. Komatsubara}

\address{for the E787 Collaboration\\KEK-IPNS, Oho 1-1, Tsukuba, Ibaraki 305-0801, 
Japan\\E-mail: takeshi.komatsubara@kek.jp}

\twocolumn[\maketitle\abstract{
Recent results 
from a rare kaon decay experiment E787 at the BNL-AGS
on \pinn, \kmu2g, and \kpi2g decays
are reported.}]

\section{\pinn}

\subsection{Theoretical Motivation}\label{subsec:theory}
The \pinn decay is a flavor changing neutral current process
induced in the Standard Model (SM) by loop effects 
in the form of penguin and box diagrams. 
The decay is sensitive to top-quark effects 
and provides an excellent route 
to determine the absolute value of $V_{td}$
in the Cabibbo-Kobayashi-Maskawa matrix.
With the constraints from other K and B decay experiments 
the SM prediction of the branching ratio
is $(0.82\pm0.32)\times 10^{-10}$, and using
only the results on $B_d-\bar{B_d}$ and $B_s-\bar{B_s}$ mixing 
a branching ratio limit $<1.67\times 10^{-10}$ can be extracted\cite{BB99}.
New physics beyond the SM could affect the branching ratio\cite{beyondSM}. 

\subsection{E787 Detector}\label{subsec:detector}
\begin{figure}
\epsfxsize120pt
\figurebox{120pt}{160pt}{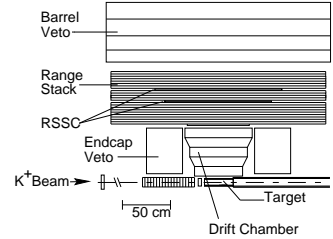}
\caption{Side view of the upper half of the E787 detector.} 
\label{fig:detector}
\end{figure}

E787\footnote{E787 is a collaboration of BNL,
 Fukui, KEK, Osaka, Princeton, TRIUMF, Alberta and British Columbia.}
measures the charged track emanating from stopped $K^+$ decays. 
The E787 detector (Figure\ref{fig:detector})  
is a solenoidal spectrometer
with a 1.0 Tesla field directed along the LESB3 beam line\cite{LESB3}.
Slowed by a BeO degrader, kaons stop in the scintillating-fiber target 
at the center of the detector. 
A delayed coincidence requirement ($>$ 2nsec)
between the stopping kaon and the outgoing pion times
helps to reject backgrounds of pions scattered into the detector
or kaons decaying in flight.
Charged decay products pass through the drift chamber, 
lose energy by ionization loss and stop 
in the Range Stack made of plastic scintillators and straw chambers.
Momentum, kinetic energy and range are measured 
to reject the backgrounds by kinematic means.
For further rejection of $\mu^+$ tracks, 
the output pulse-shapes of the Range Stack counters
are recorded and analyzed so that the decay chain 
$\pi^+\to\mu^+\to e^+$ can be identified in the stopping scintillator. 
$K^+\to\pi^+\pi^0$ 
and other decay modes with extra particles
($\gamma$, $e$, ...) are vetoed by the coincident signals 
in the hermetic shower counters. 

\subsection{Result and Prospect}\label{subsec:pnnresult}
\begin{figure}
\epsfxsize120pt
\figurebox{120pt}{160pt}{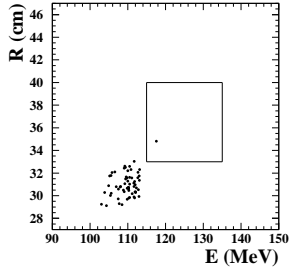}
\caption{Range vs kinetic energy plot of the final \pinn sample
         from the E787 1995-1997 data set. The box indicates 
         the signal acceptance region.}
\label{fig:pnnplot}
\end{figure}

E787 took data on \pinn from 1995 to 1998.
In the 1995 data set, 
one event 
was observed\cite{E787-95} in the signal region. 
The new result\cite{E787-9597} from the 1995-1997 data set
is shown in Figure\ref{fig:pnnplot}.
The same one event was observed
and no new events were found in the signal region. 
The new value for the branching ratio is 
$(1.5^{+3.4}_{-1.2})\times 10^{-10}$. 
Compared to the result from 1995,  
2.1 times more kaon exposure ($3.2\times 10^{12}$)
and 30\% better acceptance
are achieved. 
The background level, $0.08\pm 0.02$ events,  
corresponding to a branching ratio of 1.2$\times 10^{-11}$, 
is improved by a factor of 2.5. 
The new result provides a constraint  
$1.07\times 10^{-4} < |V_{ts}^{*}V_{td}| < 1.39\times 10^{-3}$
without reference to the B system. 

The kaon exposure in the E787 1998 data set 
is comparable to 1995-1997, and 
the analysis is ongoing. 
The sensitivity for the entire E787 data 
is expected to reach $0.7\times 10^{-10}$,  
which is comparable to the SM prediction.

E949\cite{E949}
continues the experimental study of \pinn at the AGS
based on the experience of E787
and is expected to reach a sensitivity of $10^{-11}$ or less
in two to three years of operation. 
An engineering run of E949 is scheduled for 2001.

\section{Photon Detection in E787}\label{sec:BVdet}
The Barrel Veto counter (BL) 
in Figure\ref{fig:detector}
detects photons
in the study of radiative kaon decays.
The counter consists of 48 azimuthal sectors and 4 radial layers, 
made of lead/scintillator 14 radiation lengths in depth, and 
covers a solid angle of about 3$\pi$ sr. 
The position of BL hits along the beam line is measured 
with ADC and TDC information from phototubes 
on both ends of each 2-m long module. 
The energy and direction of the photons from stopped kaon decays
are determined 
from the offline clustering in the BL and the decay 
vertex position 
in the target. 

\section{\kmu2g}\label{sec:kmu2gamma}
The decay \kmu2g can proceed via 
internal bremsstrahlung (IB)
and structure dependent decay (SD).
The latter is sensitive to the electroweak structure of 
the kaon because the photon is emitted from intermediate states.
E787 has made the first measurement\cite{kmng}
of the SD$^{+}$ component in \kmu2g
decay, which is 
proportional to the square of the absolute value 
of the sum of the Vector and Axial form factors
($|F_{V}+F_{A}|$).

\begin{figure}
\epsfxsize120pt
\figurebox{120pt}{160pt}{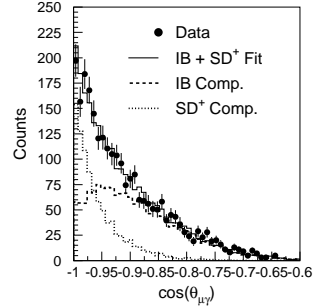}
\caption{Counts vs. $\cos(\theta_{\mu\gamma})$ of the \kmu2g events, 
         and various fits 
         as described in the text.}
\label{fig:kmu2gplot}
\end{figure}

The SD$^{+}$ component peaks at high muon and photon energy.
With a total kaon exposure of $9.2\times 10^{9}$
and $1.5\times 10^{6}$ triggers for \kmu2g, 
2693 events are observed in the signal region
where the $\mu^+$ kinetic energy is $>$ 137MeV and 
the $\gamma$ energy is $>$ 90MeV.
The distribution of the opening angle between $\mu^+$ and $\gamma$
($\cos(\theta_{\mu\gamma})$)
for background-subtracted data, shown in Figure\ref{fig:kmu2gplot}
with the Monte Carlo distributions for IB and SD$^{+}$ components 
superimposed, 
clearly indicates that the SD$^{+}$ component is present. 
Detailed fits yield
$|F_{V}+F_{A}|=0.165\pm0.007\pm0.011$, which corresponds to 
an SD$^{+}$ branching ratio of $(1.33\pm 0.12\pm 0.18)\times 10^{-5}$, and 
a 90\% confidence level limit $-0.04 < F_{V}-F_{A} < 0.24$.

\section{\kpi2g}\label{sec:kpi2gamma}
The decay \kpi2g in which the photon is directly emitted (DE) 
is sensitive to the low energy hadronic interactions of mesons. 
E787 has performed a new measurement\cite{kp2g}
of \kpi2g decay, with significantly higher statistics 
than before and improved kinematic constraints 
using the $K^+$ decays at rest.
The DE component is 
isolated kinematically with the variable $W^2$,
 which is reconstructed from the opening angle 
 between $\pi^+$ and $\gamma$ ($\cos(\theta_{\pi^+\gamma})$), 
 $\pi^+$ energy and momentum ($E_{\pi^+}$,$P_{\pi^+}$), 
 and $\gamma$ energy ($E_{\gamma}$)
 as 
 $W^2= E_{\gamma}^2 \times (E_{\pi^+}-P_{\pi^+}\times 
  \cos(\theta_{\pi^+\gamma}))/(m_{K^+} \times m_{\pi^+}^2)$
 in the stopped $K^+$ decays
 and is directly related to the observables in the E787 detector.

\begin{figure}
\epsfxsize120pt
\figurebox{120pt}{160pt}{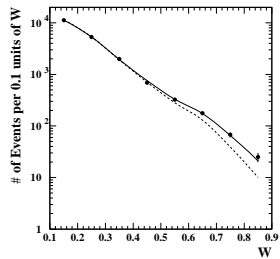}
\caption{W spectrum of the observed \kpi2g events and best fits to 
         IB+DE (solid curve) and IB alone.}
\label{fig:kpi2gplot}
\end{figure}

With a total kaon exposure of $1.8\times 10^{11}$
and $1.1\times 10^{7}$ ``3gamma'' triggers to detect the charged track
and three $\gamma$ clusters in the BL, 
19836 events survived all selection cuts
including requiring that the kinematically fitted $\pi^+$ momentum 
be between 140 and 180 MeV/$c$. 
Figure\ref{fig:kpi2gplot} shows the $W$ spectrum of the signal events.
The DE component is measured to be $(1.8\pm0.3)$\% of the IB component, 
yielding a branching ratio for DE of $(4.7\pm 0.8\pm 0.3)\times 10^{-6}$
in the $\pi^+$ kinetic energy range 55-90 MeV\footnote{
  Previous experiments used decay-in-flight techniques. 
  The current Particle Data Group average\cite{PDG00} 
  is $(1.8\pm0.4)\times 10^{-5}$.}.
This result can be understood by purely magnetic contributions
in the framework of Chiral Perturbation Theory.

\section*{Acknowledgments}
This research was supported in part 
by the U.S. Department of Energy under Contracts 
No. DE-AC02-98CH10886,  
W-7405-ENG-36, 
and grant DE-FG02-91ER40671, 
by the Ministry of Education, Science, Sports and Culture of Japan
through the Japan-U.S. Cooperative Research Program in High Energy Physics
and under the Grant-in-Aids for Scientific Research, 
for Encouragement of Young Scientists and for JSPS Fellows, 
and by the Natural Sciences and Engineering Research Council 
and the National Research Council of Canada.

\end{document}